\documentclass[aps,10pt, pra,superscriptaddress,twocolumn]{revtex4-2}

\usepackage{amsfonts,amssymb,amsmath}
\usepackage{graphics,graphicx,epsfig}

\newtheorem{theorem}{Theorem}

\newcommand{\ket}[1]{|{#1}\rangle}

\begin{document}

\title{Scalable Test of Genuine Multipartite Entanglement via Partially Randomized Measurements}

\author{Jan W{\'o}jcik}
\affiliation{Institute of Theoretical Physics and Astrophysics, University of Gda\'nsk, 80-308 Gda\'nsk, Poland}

\author{Pawe{\l} Chrabkowski}
\affiliation{Institute of Theoretical Physics and Astrophysics, University of Gda\'nsk, 80-308 Gda\'nsk, Poland}

\author{Wies{\l}aw Laskowski}
\affiliation{Institute of Theoretical Physics and Astrophysics, University of Gda\'nsk, 80-308 Gda\'nsk, Poland}

\begin{abstract}
Certifying genuine multipartite entanglement in quantum systems can require a number of measurements that grows exponentially with the system size. Here we introduce a criterion based on correlation-tensor subsector lengths restricted to local measurement planes and show that it can be evaluated using partially randomized measurements without an explicit exponential dependence on the number of qubits. We derive the corresponding bounds for $k$-separable states and illustrate the criterion using representative families of multipartite entangled states. Finally, we demonstrate the practical applicability of the method on an ion-trap quantum computer by certifying genuine five-partite entanglement.
\end{abstract}

\maketitle

\section{Introduction}

Entanglement is a fundamental resource in quantum information science, enabling advantages in quantum communication, computation, and metrology \cite{horodecki2009,Friis:2018nyl}. In multipartite systems, a particularly strong form of this resource is genuine multipartite entanglement (GME), in which quantum correlations are genuinely shared among all parties. More precisely, an $N$-partite state is genuinely multipartite entangled if it cannot be expressed as a convex mixture of states that are separable across different bipartitions.

A variety of methods for certifying GME have been developed. These include state-specific and stabilizer-based entanglement witnesses \cite{TothGuhne2005}, criteria formulated in terms of density-matrix elements \cite{GuhneSeevinck2010}, correlation tensors \cite{deVicenteHuber2011, PhysRevA.84.062305, PhysRevA.87.034301, PhysRevLett.114.180501}, semidefinite-programming characterizations based on PPT mixtures \cite{Jungnitsch2011}, device-independent approaches based on Bell-type inequalities \cite{Svetlichny1987, Collins2002,SeevinckSvetlichny2002, Toth2005, Laskowski2005,Bancal2011}, and criteria based on the quantum Fisher information \cite{PhysRevA.85.022322,PhysRevA.85.022321}. Although these methods can provide powerful and experimentally efficient tests, many require prior knowledge of the target state, carefully chosen measurement settings, or well-aligned local reference frames.

To overcome these limitations, randomized local measurements provide a complementary approach that substantially reduces the need for measurement optimization and reference-frame alignment. Since the measurement directions are sampled randomly, entanglement can be characterized without selecting specific settings or establishing a shared reference frame. In particular, statistical moments and distributions of full and marginal correlations obtained from randomized measurements yield experimentally accessible criteria for multipartite entanglement and GME \cite{Ketterer2019,Knips2020,Imai2021}.

In this work, we extend this framework to partially randomized measurements, in which local measurement directions are sampled from restricted regions. We derive a GME criterion for this scenario and establish the corresponding bounds for biseparable and, more generally, $k$-separable states. We illustrate the criterion using several representative multipartite states and demonstrate its practical applicability on a trapped-ion quantum processor.

\section{GME criteria}

Consider an $N$-qubit state $\rho$ with correlation tensor elements defined as $T_{\vec n_1 \dots \vec n_N} = {\rm Tr}(\rho \vec n_1 \cdot \vec \sigma \otimes \dots \otimes \vec n_N \cdot \vec \sigma)$, where $\vec \sigma = (\sigma_x,\sigma_y,\sigma_z)$ is the vector of Pauli matrices and $\vec n_i$ is a unit vector representing the measurement direction chosen by party $i$. In this work, we focus exclusively on the subsector of the correlation tensor defined by local planes $\mathcal{P}_i$ spanned by pairs of orthogonal vectors $\{\vec n^{1}_i,\vec n^{2}_i\}$. Each party may choose its pair independently.

The fundamental constraint on this subsector length \cite{Aschauer2004}, following from the anticommutation relations of Pauli operators \cite{kurzynski2011}, is given by
\begin{equation}
\|T\|_{\mathcal{P}_1...\mathcal{P}_N}^2=\sum_{i_1, \dots, i_N=1}^2 T_{\vec n_1^{i_1} \dots \vec n_N^{i_N}}^2 \leq 2^{N-1}.
\label{ar}
\end{equation}

We first establish a bound on the subsector length for $k$-product states. A state $\rho$ is $k$-product if it factorizes into $k$ independent subsystems of sizes $\{l^u\}_{u=1}^k$, such that $\sum_{u=1}^k l^u = N$.

Since the correlation tensor of a $k$-product state factorizes as
\begin{equation}
T_{i_1...i_N} = T_{i_1...i_{l^1}}^{(1)} \cdots T_{i_1...i_{l^k}}^{(k)} = \prod_{u=1}^k T^{(u)}_{i_1...i_{l^u}},
\end{equation}
where $T^{(u)}$ denotes the correlation tensor of the $u$-th constituent subsystem, the subsector length factorizes accordingly:
\begin{eqnarray}
\|T\|^2_{\mathcal{P}_1...\mathcal{P}_N}&=&\sum_{i_1, \dots, i_N=1}^2 T_{\vec n_1^{i_1} \dots \vec n_N^{i_N}}^2 \\
&=& \sum_{i_1, \dots, i_N=1}^2 \prod_{u=1}^k (T_{\vec n_1^{i_1} \dots \vec n_{l^u}^{i_{l^u}}}^{(u)})^2 \nonumber \\
&=& \prod_{u=1}^k \sum_{i_1, \dots, i_{l^u}=1}^2 (T_{\vec n_1^{i_1} \dots \vec n_{l^u}^{i_{l^u}}}^{(u)})^2. \label{eq}
\end{eqnarray}

Applying the bound (\ref{ar}) to each constituent subsystem,
$\sum_{i_1, \dots, i_{l^u}=1}^2 (T_{\vec n_1^{i_1} \dots \vec n_{l^u}^{i_{l^u}}}^{(u)})^2 \leq 2^{l^u-1}$, we obtain
\begin{eqnarray}
\|T\|^2_{\mathcal{P}_1 ...\mathcal{P}_N}&\leq& \prod_{u=1}^k 2^{l^u-1} \nonumber \\
&=& 2^{\sum_{u=1}^k (l^u -1)} = 2^{N-k}.
\end{eqnarray}

To extend this result to the broader class of $k$-separable states, consider a convex mixture $\rho_{k\text{-sep}} = \sum_m p_m \rho^{(m)}$, where each $\rho^{(m)}$ is a $k$-product state, possibly with respect to a different partition. Using the linearity of the correlation tensor and the convexity of the squared norm, we obtain
\begin{eqnarray}
\label{sep}
\|T\|^2_{\mathcal{P}_1...\mathcal{P}_N} &=& \sum_{i_1, \dots, i_N=1}^2 \left( \sum_m p_m T_{\vec n_1^{i_1} \dots \vec n_N^{i_N}}^{(m)} \right)^2 \nonumber \\
&\leq& \sum_{i_1, \dots, i_N=1}^2 \sum_m p_m \left( T_{\vec n_1^{i_1} \dots \vec n_N^{i_N}}^{(m)} \right)^2 \nonumber \\
&=& \sum_m p_m \left[ \sum_{i_1 \dots i_N=1}^2 \left( T_{\vec n_1^{i_1} \dots \vec n_N^{i_N}}^{(m)} \right)^2 \right] \nonumber \\
&\leq& \sum_m p_m 2^{N-k} = 2^{N-k}.
\end{eqnarray}

These results show that $2^{N-k}$ is a universal upper bound on the subsector length for $k$-separable states. A violation of this bound therefore excludes $k$-separability. This leads to the following theorem.

\begin{theorem}
If the subsector length of the correlation tensor of a state $\rho$, defined with respect to the local planes $\mathcal{P}_i$ $(i=1,\ldots,N)$, exceeds the threshold value $2^{N-k}$, i.e.,
\begin{equation}\label{th1}
\|T\|^2_{\mathcal{P}_1 \dots \mathcal{P}_N} > 2^{N-k},
\end{equation}
then the state $\rho$ is not $k$-separable. In particular, for $k=2$, the criterion certifies genuine $N$-partite entanglement.
\end{theorem}

\subsection{Randomized measurements}

Importantly, the subsector length is directly related to the average squared correlation, as previously shown \cite{Seymour1984,tran2015,tran2016, Ketterer2019,Wyderka2020}. For the subsector length, this relation takes the following form
\begin{equation}\label{tdesign}
\|T\|^2_{\mathcal{P}_1 \dots \mathcal{P}_N} = 2^N  \mathcal{R}^{(2)}_{\mathcal{P}_1 \dots \mathcal{P}_N},
\end{equation}
where the average squared correlation is given by
\begin{equation}
\mathcal{R}^{(2)}_{\mathcal{P}_1 \dots \mathcal{P}_N}= \frac{1}{(2\pi)^N} \int_0^{2\pi} d\phi_1 \dots  d\phi_N E_{\mathcal{P}_1 \dots \mathcal{P}_N}(\phi_1,\dots,\phi_N)^2,
\label{int}
\end{equation}
where
\begin{eqnarray}
&&E_{\mathcal{P}_1 \dots \mathcal{P}_N}(\phi_1,\dots,\phi_N) = \nonumber\\
&& {\rm Tr}[ \rho (\cos \phi_1 \vec n^1_1 \cdot \vec \sigma + \sin \phi_1 \vec n^2_1 \cdot \vec \sigma) \otimes ...  \label{ekor}\\
&& ... \otimes (\cos \phi_N \vec n^1_N \cdot \vec \sigma + \sin \phi_N \vec n^2_N \cdot \vec \sigma)] \nonumber
\end{eqnarray}
is the correlation function for measurement directions in the planes $\mathcal{P}_i$, specified by the angles $\phi_i$ ($i=1,...,N$). Combining (\ref{sep}) and (\ref{tdesign}) leads to the following theorem.

\begin{theorem}
If the average squared correlation $\mathcal{R}_{\mathcal{P}_1 \dots \mathcal{P}_N}^{(2)}$ exceeds the threshold value $2^{-k}$, i.e.,
\begin{equation}\label{th2}
\mathcal{R}_{\mathcal{P}_1 \dots \mathcal{P}_N}^{(2)}>2^{-k},
\end{equation}
then the state is not $k$-separable. In particular, for $k=2$, the criterion certifies genuine $N$-partite entanglement.
\end{theorem}

The significance of these bounds lies in the direct connection between the average squared correlation $\mathcal{R}^{(2)}$ and randomized measurements \cite{tran2015}. The average squared correlation can be estimated experimentally by approximating the integral in (\ref{int}) with the average of the squared correlation function given in (\ref{ekor}), evaluated for angles $\phi_i$ sampled uniformly from $[0,2\pi)$. Note that estimating $\mathcal{R}_{\mathcal{P}_1 \dots \mathcal{P}_N}^{(2)}$ does not introduce any explicit dependence on the number of parties $N$ and can be performed using polynomial resources, making the method scalable to multipartite quantum systems.

Notably, the bounds for excluding successive levels of $k$-separability form a ladder-like structure, with consecutive thresholds differing by a factor of $1/2$. Specifically, $\mathcal{R}_{\mathcal{P}_1 \dots \mathcal{P}_N}^{(2)}>1/4$ excludes biseparability, while $\mathcal{R}_{\mathcal{P}_1 \dots \mathcal{P}_N}^{(2)}>1/8$ excludes $3$-separability, and so forth.

We note that Ref. \cite{Laskowski2005} introduced a criterion for genuine $N$-partite entanglement that can be adapted to yield the criterion given in (\ref{th1}) and presented in Theorem 1. However, its derivation relies on a different approach based on Bell inequalities, and the use of randomized measurements was not discussed there.

\section{Examples}

We now examine the performance of our criterion for several well-known families of multipartite entangled states, chosen to represent qualitatively different types of correlation structures. In all examples considered below, the local planes are identical, $\mathcal{P}_1 = \dots = \mathcal{P}_N = \mathcal{P}$, and we therefore use the shorthand notation $\mathcal{R}^{(2)}_{\mathcal{P}}$ and $\|T\|_{\mathcal{P}}^2$.

\subsection{$N$-qubit GHZ state}

For the $N$-qubit GHZ state, the optimal local planes are spanned by the $x$ and $y$ directions. Exactly half of the correlation-tensor elements in the $xy$-subsector are nonzero, each equal to $\pm1$, giving 
\begin{equation}\label{r2}
    \mathcal{R}_{xy}^{(2)} = \frac{\|T\|^2_{xy}}{2^N} =  \frac{\sum_{i_1, \dots, i_N=\{ x,y\}} T_{i_1...i_N}^2}{2^N} = 1/2.
\end{equation}
Since $\mathcal{R}^{(2)}_{xy} = 1/2 > 2^{-k}$ already at $k=2$, genuine $N$-partite entanglement is certified for every $N$. 

\subsection{$N$-qubit Dicke states}

We next consider the $N$-qubit Dicke states \cite{Dicke1954},
\begin{equation}
\ket{D^{(e)}_{N}} = \binom{N}{e}^{-\frac{1}{2}}\sum_{k}{\mathcal{P}_{k}(\ket{0}^{\otimes{(N-e)}}\otimes{\ket{1}^{\otimes{e}}})},
\end{equation}
where $\sum_k\mathcal{P}_k(\cdot)$ denotes the sum over distinct permutations of zeros and ones. For this family, the subsector length is maximal when all local planes are chosen to be the $xz$ plane, yielding
\begin{equation}
    \mathcal{R}_{xz}^{(2)}=\frac{\|T\|^2_{xz}}{2^N}=\frac{1}{2^N}\sum_{m=0}^{\min(e, N-e)} \frac{\binom{e}{m}^2\binom{N-e}{m}^2}{\binom{N}{2m}}.
    \label{dsum}
\end{equation}
A derivation is provided in Appendix \ref{appendixA}. Figure~\ref{fig:wykres} shows $\mathcal{R}_{xz}^{(2)}$ for representative $N \leq 10$. With the genuine-entanglement threshold $\mathcal{R}_{xz}^{(2)} > 2^{-2}=0.25$, our criterion certifies $D^{(1)}_{3}$ and $D^{(2)}_{4}$ as genuinely $N$-partite entangled, and rules out $3$-separability for $D^{(1)}_{4}$, $D^{(2)}_{5}$, $D^{(3)}_{6}$, $D^{(3)}_{7}$, and $D^{(4)}_{8}$.

\begin{figure}
    \centering
    \includegraphics[width=0.95\linewidth]{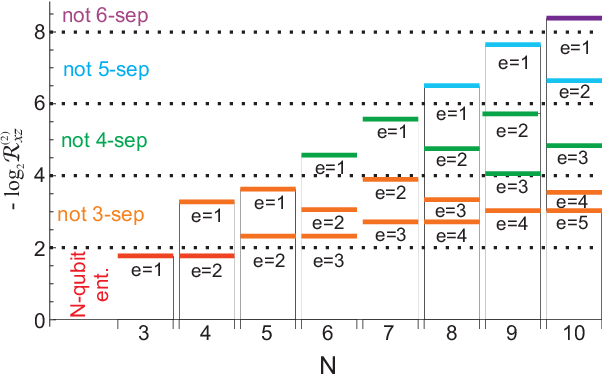}
    \caption{The value of $-\log_2\mathcal{R}^{(2)}_{xz}$ for different Dicke states $\ket{D^{(e)}_{N}}$, with the highlighted intervals corresponding to the entanglement levels detected by our criterion.}
    \label{fig:wykres}
\end{figure}

\subsection{Family of four-qubit states}

Finally, we consider the four-qubit family $\ket{\psi(\alpha)}$ \cite{Wieczorek_2008,Schmid_2008phd},
\begin{eqnarray}
\ket{\psi(\alpha)}&=&\sqrt{\alpha}\left(\frac{1}{\sqrt{2}}(\ket{0011}+\ket{1100})\right)\\
&+&\sqrt{1-\alpha}\left(\frac{1}{2}(\ket{01}+\ket{10})(\ket{01}+\ket{10})\right).\nonumber
\end{eqnarray}

For this family, we consider the $xy$ and $xz$ local planes. The corresponding average squared correlations are
\begin{equation}
\mathcal{R}^{(2)}_{xy} = \frac{1}{4}\left(1 + \alpha(3\alpha-2)\right),
\end{equation}
and
\begin{equation}
\mathcal{R}^{(2)}_{xz} = \frac{1}{8}\left(2-\alpha(3\alpha-2)\right),
\end{equation}
respectively. For each value of $\alpha$, the stronger criterion is obtained by taking the larger of these two quantities:
\begin{equation}
\max_{\vec n^1 \vec n^2} \mathcal{R}_{\vec n^1 \vec n^2}^{(2)} = \left \{
\begin{array}{ll}
\mathcal{R}^{(2)}_{xz} & {\rm for~}\alpha \leq 2/3\\
\mathcal{R}^{(2)}_{xy} & {\rm for~}\alpha \geq2/3.
\end{array} \right.
\label{maks}
\end{equation}

\begin{figure}
\centering
\includegraphics[width=0.95\linewidth]{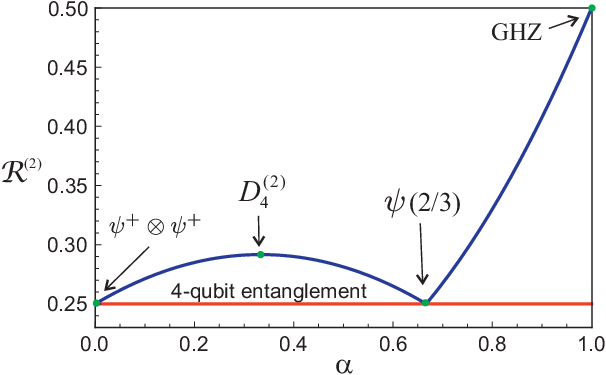}
\caption{The maximal value of $\mathcal{R}^{(2)}$ (\ref{maks}) as a function of $\alpha$, with the red line indicating the threshold above which our criterion certifies genuine four-partite entanglement. The marked points $\alpha = 0$, $\alpha = 1/3$, $\alpha=2/3$, and $\alpha = 1$ correspond to the states $\ket{\psi^{+}}\otimes\ket{\psi^+}$, the Dicke state $D^{(2)}_{4}$, $\psi(2/3)$, and the GHZ state, respectively.}
\label{fig:alpha}
\end{figure}

Genuine multipartite entanglement is certified for every $\alpha\in[0,1]$ except at $\alpha=0$ and $\alpha=2/3$, where the maximal value is exactly $1/4$. Since the criterion requires a strict violation of this threshold, these two cases are not detected (see Fig.~\ref{fig:alpha}).

The two boundary points have different physical origins. At $\alpha=0$, $\ket{\psi(\alpha)}$ reduces to $\ket{\psi^+}\otimes\ket{\psi^+}$, a product of two Bell pairs across the $12|34$ partition. The state is therefore biseparable and not genuinely four-partite entangled. The situation at $\alpha=2/3$ is qualitatively different: $|\psi(2/3)\rangle$ is genuinely four-partite entangled, but lies exactly at the detection threshold and is therefore not detected by our criterion. This illustrates a general feature of entanglement witnesses: the criterion is sufficient but not necessary, and states saturating the bound need not be detected.

\section{Performance}

We now compare the statistical performance of the direct subsector-length and randomized-measurement approaches. The aim is to estimate how the corresponding uncertainties scale with the number of qubits \(N\) and with the total number of measurements.

We first consider the subsector length criterion (\ref{th1}).
Each correlation $T_{\vec n_1\ldots \vec n_N}$ is estimated from $n$ shots. Since the outcomes are dichotomic, the standard deviation of a single correlation estimate is
\begin{equation}
    \Delta T_{\vec n_1\ldots \vec n_N}
    =
    \sqrt{\frac{1-T_{\vec n_1\ldots \vec n_N}^2}{n}} .
\end{equation}
Therefore, by standard error propagation,
\begin{eqnarray}
    \Delta\left(T_{\vec n_1\ldots \vec n_N}^2\right)
    &=&
    2|T_{\vec n_1\ldots \vec n_N}|\Delta T_{\vec n_1\ldots \vec n_N}
    \\
   &=& \frac{
    2|T_{\vec n_1\ldots \vec n_N}|
    \sqrt{1-T_{\vec n_1\ldots \vec n_N}^2}
    }{\sqrt n}.\nonumber
\end{eqnarray}
The propagated uncertainty of the full subsector length criterion is then
\begin{equation}
    \Delta^2\left(\|T\|_{\mathcal{P}_1 \dots \mathcal{P}_N}^2\right)
    =
    \frac{1}{n}
    \sum_{i_1, \dots, i_N=1}^2
    4T_{\vec n_1^{i_1}\ldots \vec n_N^{i_N}}^2
    \left(1-T_{\vec n_1^{i_1}\ldots \vec n_N^{i_N}}^2\right).
\end{equation}
Using the bound
\begin{equation}
    4T^2(1-T^2)\leq 1,
\end{equation}
we obtain
\begin{equation}
    \Delta^2\left(\|T\|_{\mathcal{P}_1 \dots \mathcal{P}_N}^2\right)
    \leq
    \frac{2^N}{n}.
\end{equation}
Thus, to guarantee a target uncertainty
\(\Delta(\|T\|_{\mathcal{P}_1 \dots \mathcal{P}_N}^2)\leq \epsilon\), it is sufficient to take
\begin{equation}
    n \geq \frac{2^N}{\epsilon^2}
\end{equation}
shots per correlation term. Since the subsector length criterion contains
\(2^N\) correlation terms, the total number of measurements scales as
\begin{equation}
    M_{\|T\|_{\mathcal{P}_1 \dots \mathcal{P}_N}^2}
    =
    2^N n
    \sim \frac{4^N}{\epsilon^2}.
\end{equation}

We now turn to the randomized criterion  (\ref{th2}). In this case one estimates the averaged quantity $\mathcal{R}_{\mathcal{P}_1 \dots \mathcal{P}_N}^{(2)}$
over $m$ randomly chosen measurement settings. Each correlation $E^{(j)}_{\mathcal{P}_1 \dots \mathcal{P}_N} \equiv E^{(j)}_{\mathcal{P}_1 \dots \mathcal{P}_N}(\phi_1^{(j)}, ..., \phi_N^{(j)})$ with $j=1,...,m$ is estimated from $n'$ shots. The uncertainty of \(\mathcal{R}_{\mathcal{P}_1 \dots \mathcal{P}_N}^{(2)}\) has two contributions. The first one is the statistical uncertainty associated with estimating each correlation from a finite number of shots. The second one is the sampling uncertainty resulting from averaging over only \(m\) randomly drawn settings.  Adding the corresponding variances gives
\begin{eqnarray}
&&\Delta^2 \mathcal{R}_{\mathcal{P}_1 \dots \mathcal{P}_N}^{(2)}
= \frac{1}{m^2 n'} \sum_{j=1}^m
4(E^{(j)}_{\mathcal{P}_1 \dots \mathcal{P}_N})^2\left(1-(E^{(j)}_{\mathcal{P}_1 \dots \mathcal{P}_N})^2\right) \nonumber \\
&&+ \frac{1}{m(m-1)} \sum_{j=1}^m
\left((E^{(j)}_{\mathcal{P}_1 \dots \mathcal{P}_N})^2-\mathcal{R}^{(2)}_{\mathcal{P}_1 \dots \mathcal{P}_N}\right)^2 .
\end{eqnarray}
Using again \(4(E^{(j)}_{\mathcal{P}_1 \dots \mathcal{P}_N})^2(1-(E^{(j)}_{\mathcal{P}_1 \dots \mathcal{P}_N})^2)\leq 1\), the first term is bounded by
\begin{equation}
    \frac{1}{m^2 n'}
    \sum_{j=1}^m 4(E^{(j)}_{\mathcal{P}_1 \dots \mathcal{P}_N})^2\left(1-(E^{(j)}_{\mathcal{P}_1 \dots \mathcal{P}_N})^2\right)
    \leq \frac{1}{m n'} .
\end{equation}
Moreover, since $0\leq (E^{(j)}_{\mathcal{P}_1 \dots \mathcal{P}_N})^2\leq 1$, the sampling variance is bounded by
\begin{equation}
    \mathrm{Var}((E^{(j)}_{\mathcal{P}_1 \dots \mathcal{P}_N})^2)\leq \frac14 .
\end{equation}
Consequently,
\begin{equation}
    \Delta^2 \mathcal{R}_{\mathcal{P}_1 \dots \mathcal{P}_N}^{(2)}
    \leq
    \frac{1}{m n'}
    +
    \frac{1}{4m}.
\end{equation}
This expression contains no explicit factor growing exponentially with the number of qubits. To reach a target uncertainty
\(\Delta \mathcal{R}_{\mathcal{P}_1 \dots \mathcal{P}_N}^{(2)}\leq \epsilon\), it is sufficient to take
\begin{equation}
    m \sim \frac{1}{\epsilon^2}
\end{equation}
random settings and a constant number of shots per setting. Therefore, the total number of measurements in the randomized scheme scales as
\begin{equation}
    M_{\mathcal{R}_{\mathcal{P}_1 \dots \mathcal{P}_N}^{(2)}}
    = m n' \sim \frac{1}{\epsilon^2}.
\end{equation}

This comparison shows that the subsector-length criterion incurs a measurement overhead scaling as $4^N$, whereas the randomized criterion requires a number of measurements scaling as $1/\epsilon^2$, with no explicit dependence on $N$. This establishes the statistical advantage of the latter.

\section{Demonstration}

We accompany the theoretical results with a demonstration made on the PIAST-Q trapped-ion quantum computer. We have used a 5-qubit circuit in order to show the genuine entanglement of 5-qubit GHZ state using the two presented approaches, i.e., the standard subsector-length approach measuring all $2^{N-1}=16$ nonzero correlation tensor elements and using uniform randomized measurements in the $xy$ equatorial plane to obtain the $\mathcal{R}_{xy}^{(2)}$.

\begin{figure}
    \centering
    \includegraphics[width=0.95\linewidth]{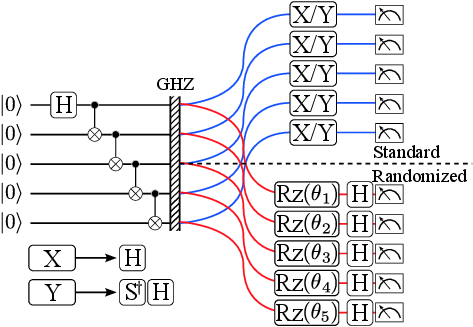}
    \caption{Scheme of the circuit. First, Hadamard and CNOT gates are applied to prepare the GHZ state. The standard subsector-length approach of (\ref{th1}) and the randomized-measurement approach of (\ref{th2}) are then implemented, as indicated by the blue and red branches, respectively. The additional gates $S^\dagger = (\sqrt{Z})^\dagger$ and $R_z(\theta) = e^{-i\frac{\theta}{2}\sigma_z}$ are used to perform the required measurements.}
    \label{fig:circuit}
\end{figure}

The circuit scheme is shown in Fig.~\ref{fig:circuit}. First, the GHZ state is prepared, after which the two measurement approaches are implemented, as indicated by the blue and red branches in the figure. In the standard approach, each qubit is measured along either the $x$ or $y$ direction using the Hadamard gate and the $S^\dagger$ gate, where $S^\dagger = (\sqrt{Z})^\dagger$. In the randomized approach instead of measuring only in $xy$ directions, we measure uniformly distributed directions from the entire $xy$ plane. To this end we use the Hadamard and the $R_z(\theta)$ gates which describe the rotations of the qubit on the Bloch sphere around the $z$ axis, i.e., $R_z(\theta) = e^{-i\frac{\theta}{2}\sigma_z}$ on each qubit with randomly chosen $\theta$. 

In both approaches we have used 200 shots per measurement setting. In the subsector-length approach, we obtained $\|T\|^2_{xy} = 8.55 \pm 0.07$, which exceeds the bound $2^{5-2} = 8$ and thus certifies genuine five-partite entanglement.

In the random-measurement approach, we obtained $\mathcal{R}_{xy}^{(2)} = 0.27775 \pm 0.00998$, which exceeds the bound $2^{-k} = 1/4$ and again certifies genuine five-partite entanglement. Both results confirm the genuine five-partite entanglement of the prepared GHZ state.

\section{Conclusions}
We have presented a criterion for detecting genuine multipartite entanglement based on correlation-tensor subsector lengths defined within local measurement planes. We have shown that the criterion can be evaluated using randomized measurements performed within these planes. The resulting bounds do not depend explicitly on the system size, in contrast to the direct subsector-length approach, whose measurement cost grows exponentially with the number of qubits.

We have illustrated the performance of the criterion for representative families of multipartite entangled states and complemented the analytical results with a demonstration on an ion-trap quantum computer. The experiment confirms that the method can be used to certify genuine multipartite entanglement in a realistic noisy quantum-processing platform.

\section{Acknowledgments}
The authors are supported by the National Science Centre (NCN, Poland) within the OPUS project (Grant No. 2024/53/B/ST2/04103). We acknowledge the EuroHPC JU, the Ministry of Higher Education, and the Ministry of Digital Affairs in Poland for awarding the project with ID 5/2026 access to the EuroHPC allocation on the PIAST-Q trapped-ion quantum computer hosted by the Pozna\'n Supercomputing and Networking Center.

\section{Data Availability Statement}

The data that support the findings of this article are publicly available \cite{dataset}.

\appendix

\section{Derivation of Eq. (\ref{dsum})}

\label{appendixA}

By permutation invariance of the Dicke state $|D_N^{(e)}\rangle$, we divide the qubits into two sets: $A$ ($n_x$ qubits on which $\sigma_x$ acts) and $B$ ($n_z=N-n_x$ qubits on which $\sigma_z$ acts).

For a basis state
\begin{equation}
|s\rangle = |\underbrace{1...1}_{s}\underbrace{0...0}_{N-s}\rangle,
\end{equation}
let $s$ denote the number of qubits in the state $|1\rangle$, i.e., the number of excitations. The Dicke state $|D_N^{(e)}\rangle$ is the equal superposition of all basis states $|s\rangle$ with exactly $e$ excitations. Note that $\sigma_x$ flips $|0\rangle\leftrightarrow|1\rangle$ without introducing a phase, whereas $\sigma_z$ leaves the qubit unchanged but contributes a factor $-1$ for each excited qubit. Therefore,
\begin{equation}
\sigma_x^{\otimes n_x}\sigma_z^{\otimes n_z}|s\rangle=(-1)^{s_B}|s'\rangle,
\end{equation}
where $s_B$ is the number of excitations of $s$ in $B$, and $|s'\rangle$ is the basis state identical to $|s\rangle$ on $B$ and complementary to it on $A$, with every qubit in $A$ flipped.

The overlap $\langle D_N^{(e)}|s'\rangle$ is nonzero only if $|s'\rangle$ also has exactly $e$ excitations. If $|s\rangle$ has $s_A$ excitations in $A$, flipping all qubits in $A$ gives $|s'\rangle$ a total of $e+n_x-2s_A$ excitations. Requiring this number to equal $e$ gives $s_A=\frac{n_x}{2}$.
Consequently, $T_{i_1...i_N}=0$ whenever $n_x$ is odd.

For even $n_x$, the number of excitations in $A$ is fixed to $n_x/2$. The remaining
$e-\frac{n_x}{2}$
excitations must therefore be in $B$, so that
$s_B=e-\frac{n_x}{2}$.
Hence, all nonzero contributions have the same sign, $(-1)^{e-n_x/2}$.

The number of contributing basis states is
\begin{equation}
\binom{n_x}{n_x/2}\binom{n_z}{e-n_x/2},
\end{equation}
where the first factor counts the possible placements of excitations in $A$, and the second counts those in $B$. Each contribution is weighted by $\binom{N}{e}^{-1}$ due to the normalization of the Dicke state. Therefore,
\begin{equation}
T_{x^{\otimes n_x}z^{\otimes n_z}}=\frac{(-1)^{e-n_x/2}}{\binom{N}{e}}
\binom{n_x}{n_x/2}\binom{n_z}{e-n_x/2},
\end{equation}
with the correlation vanishing for odd $n_x$.

The $xz$-subsector length can then be written as
\begin{equation}
\|T\|^2_{xz}=\sum_{k=0}^{\lfloor N/2\rfloor}
\binom{N}{2k}\cdot
\frac{1}{\binom{N}{e}^{2}} \binom{2k}{k}^{2}\binom{N-2k}{e-k}^{2}.
\end{equation}
After straightforward algebraic transformations, this expression reduces to Eq. (\ref{dsum}).

\bibliographystyle{apsrev}
\bibliography{ref}

\end{document}